# A Unified Theory for Transient Synchronization Stability Analysis of Renewable Dominated Power Systems

Meng Zhan[1], Miao Han[1], Yayao Zhang[1], Hongsheng Xu[1], Jiabing Hu[1], Shijie Cheng[1], Jürgen Kurths[2]

[1]. State Key Laboratory of Advanced Electromagnetic Technology, School of Electrical and Electronic Engineering, Huazhong University of Science and Technology, Wuhan 430074, China.

[2]. Institute of Physics, Humboldt University Berlin, Berlin 10099, Germany.

Corresponding author:
M. Zhan, Email: zhanmeng@hust.edu.cn

## Abstract

**The change of electric power generation – from synchronous generator (SG) to converter - is generally regarded as the second revolution of power system. Different from rotor swing of SG in traditional grids mainly described by the swing equation (SE), the converter dynamics plays an indispensable role in modern renewable dominated power systems (RDPS). The high complexity of the RDPS, including spatial large-scale, nonlinearity, multi-time-scale, and even sequential switching, prevents us from fully understanding its dynamics and assessing its transient stability under large disturbance. Here, a variety of transient switching mechanism models of renewable devices relying on wind or solar energies under low-voltage ride-through are established and unified, which can be perfectly described by a generalized swing equation (GSE) under parameter changes for switching dynamics. The GSE focusing on the dominant phase-locking loop dynamics is similar to the SE. Mainly relying on the mechanical equivalence and the energy conservative principle, a substantially improved equal-area criterion method is proposed. Based on this method, even for large-scale renewable fields, the calculation errors for the critical clearing time are only about 1%. This elegant nonlinear-dynamics-based approach establishes a unified theory including modelling and analysis for the RDPS transient dynamics.**

Very recently, the penetration of renewable energy equipment into power grids, mainly including permanent-magnet synchronous generator (PMSG) and doubly-fed induction generator (DFIG) for wind power[1,2], and photovoltaic (PV) for solar power[3,4], has been continuously increasing. By the end of 2024, the total installed capacity of renewable energy generation in China has risen to around 40%[5], and that in Germany has become over 50%. Figure 1a schematically shows a renewable dominated power system (RDPS). Like the traditional energy, renewable devices should capture and transfer the primary power to electrical power and send to the AC power grid. However, different from synchronous generators (SGs)[6,7], which have been widely used in the traditional highly-controlled thermal power, hydropower, nuclear power, etc., all renewable devices interface with the power grid by power electronic converters[8], to deal with the intrinsic fluctuation of renewable energies. This change of electric power generation - from SG to converter - has fundamentally changed the power system dynamics and further all aspects of power system, including analysis[9-11], protection[12,13], control[14,15], and operation[16,17]. This is generally regarded as the second revolution of power system! Although there are already lots of engineering projects on renewable energies globally[18,19], regarded as a typical complex system[20-25], the basic theory of the RDPS remains to be established[26-28]. The RDPS stability is a great global challenge. We are still at an early stage to fully understand the organization rules of the RDPS.

Taking a closer look at any renewable energy field, such as a wind farm[29] in Fig. 1b, finds that usually it consists of several hundreds of wind turbines. They typically have radical[30], radical loop[30], or star connection[31], to feed energy into the grid by high-voltage AC (or HVDC) transmission system. Compared to the SG, they are distributed in space to make full use of the low energy density of wind or solar energy. To fully understand this spatially-distributed complex system[24], we have to investigate the renewable device first. As an example, in a single PMSG infinite-bus system[1] in Fig. 1c, it is composed of two parts: a machine-side converter (MSC) and a grid-side converter (GSC), which are connected by a DC-link capacitor. The GSC mainly includes the DC voltage control (DVC) and terminal voltage control (TVC), for controlling the active current $i_d$ and reactive current $i_q$, respectively. The MSC includes the pitch angle control, maximum power point tracking control, additional inertia control (AIC), rotor speed control, and AC current control (ACC). The AIC is indicated by a dashed box. The GSC adopts a *dq* rotating coordinate system control strategy based on the terminal voltage orientation, by the classical phase-locking loop (PLL) technique[8,32]. The PLL control diagram is shown in Fig. 1d and the coordinate transform relation is shown in Fig. 1e. By the Park transformation, the three-phase AC quantities are converted into DC quantities, so that the regulation objects are DC components. Specifically, the DVC and TVC generate the *dq*-axis current references $i_d^*$ and $i_q^*$, respectively. The ACC generates the internal potentials $e_d$ and $e_q$ by regulating the *dq* components of the output current. Finally, by the pulse width modulation technology[8,15], the six insulated gate bipolar transistor converters are driven by the modulation signal of the reference voltage $e_d$ and $e_q$, thereby generating the converter output voltage $e_{abc}$.

Apparently, the three major renewable machines including the PMSG, DFIG, and PV are exquisitely designed and optimized. Among them, the PMSG uses the SG technique and its (three-phase) stator is directly connected with the MSC and further the GSC. However, the DFIG uses the asynchronous (induction) generator technique. The stator of its induction generator is directly connected with the grid, while its rotor is connected through the MSC and GSC. In addition, similar to the PMSG, a single-stage PV is directly connected with the MSC and GSC. Clearly different from the PMSG and DFIG for the wind energy, the PV for the solar energy has no rotation component. Except these, their GSCs are nearly the same. As a result, the three major renewable devices for either wind or solar energy show different topological structures and control forms. All these make the RDPS dynamics analysis very difficult. On the other hand, it is well-known that the RDPS shows a multi-time-scale characteristic, including the ACC (around 10 ms), DVC (around 100 ms), and rotor (around 1 s) timescales. The first two belong to the electromagnetic timescale and the final one to the electromechanical timescale. In the following model simplification, we will mainly take advantage of this point.

In addition, any grid-tied renewable device should obey the compulsory requirement of low voltage ride through (LVRT) to keep connected with the grid[8]. According to the grid codes in China in Fig. 2a[33], when the positive sequence voltage at the point of common coupling of the PMSG or DFIG (PV), $U_t$, is lower than 80% (90%) of the nominal voltage, the transient switching control needs to be switched on, and the system is not allowed to offline above the red curve in Fig. 2a. Although different countries may have different grid codes[34,35], they are similar. Therefore, for the RDPS transient dynamics, the whole sequential switching process of the LVRT includes the four stages: pre-fault (stage 1), during-fault (stage 2), early fault recovery (stage 3), and late fault recovery (stage 4), which will be studied in detail.

## Unified model

1) As the first step, the four-stage mechanism models for all four typical systems are established, including the PMSG without AIC, PMSG with AIC, DFIG, and PV systems. In particular, for the PMSG without AIC, as the MSC has no direct connection with the grid[36,37], and the MSC and the GSC are completely separated, its dynamics can be regarded as dominant by the GSC. Therefore, the system is completely identical to a converter grid-tied system. Under this situation, the input power on the DC capacitor ($P_m$) is set as a constant. In a contrast, for the PMSG with AIC, the MSC and the GSC are coupled, and the input power on the SG of the PMSG ($P_m$) is set as a constant. Without losing generality, the subscripts 1-4 will be used for the variables or parameters on the corresponding stages.

The key idea in the modelling is to keep the slowest timescale dynamics and ignore the fast timescale dynamics, based on the timescale decomposition rule[38]. As the PLL dynamics for synchronization is important, it is always kept. Usually, a stable operation on stage 1 is always assumed. On stage 2, under a sudden terminal voltage dip, the device should switch to the transient switch control, as shown the LVRT box and chopper in Fig. 1c. In this situation, the PLL dynamics is dominant. Considering the reactive power support on this stage, the reactive current $i_q$ should be determined by the voltage drop depth and considered as fixed. In addition, the active current $i_d$ can be regarded as a freely chosen parameter, although together with $i_q$, it is restricted by the converter capacity limit. Next on stage 3, after the fault is cleared, $i_d$ starts to linearly increase by the active current climbing control, and the TVC is restored. Until the active current fully restores, the stage 4 starts and the system recovers completely. Hence, the models on stages 4 and 1 are the same.

The four-stage mechanism models for all four typical systems are summarized in Table 1, and more details derived from the detailed electromagnetic transient (EMT) models can be found in Supplementary Note I: Four-stage models of renewable grid-tied systems. Generally, the differential-algebraic-equation (DAE) is used to describe each stage dynamics separately. These four-stage switching models have been greatly simplified. In all four cases, their stage-1 (and stage-4) models are completely different, but their stage-2 and stage-3 models are the same. In addition, in modeling the DFIG system, as the control variable is the rotor current, by using the four scaled parameters ($a$, $b$, $c$, and $d$) to make a direct connection between the rotor current and the terminal voltage, the DFIG stage-2 and stage-3 models are similar to those of the other cases [see Supplementary Eqs. (S43)].

For model verification, these models are widely compared with the standard EMT models. Figures 2b-2h clearly show that they have caught the bulk behaviors under a voltage dip, with only some high-frequency dynamics lost. Obviously, all renewable grid-tied systems have to jump from different-order dynamical systems, based on the switch conditions. According to the nonlinear system theory, the (transient) stability should be judged after all four stages, namely, if the system can regain a stable equilibrium point at stage 4, or if the initial state of stage 4 is within the basin of attraction of the equilibrium point of stage 4[39]. Nevertheless, usually this conditional switching dynamics is very hard to analyze[40,41].

2) Next for the system analysis, although the stage-1 DAE dynamics is high-order and diversified for different cases, it only provides an initial stable condition. Its own dynamics is unimportant for the transient stability analysis. We start from the stage-2 dynamics, which is fully caught by the PLL dynamics in Fig. 1d and the terminal-voltage algebraic relation in Fig. 1e:

$$\begin{cases} \dot{\varphi} = \omega \\ \dot{\omega} = k_{ppll}\dot{u}_{tq} + k_{ipll}u_{tq} \\ u_{tq} = i_d X_g - U_g \sin\varphi \end{cases} \tag{1}$$

Here $\omega = \dot{\varphi} = \omega_{pll} - \omega_0$, and $\varphi$ represents the angle mismatch between the $d$-axis and the infinite-bus voltage phasor $\vec{U}_g$. Since the GSC adopts the $dq$ rotating coordinate control strategy based on the terminal voltage $\vec{U}_t$, in the stable operation, $\vec{U}_t$ is in phase with the $d$-axis, $u_{tq} = 0$, $\omega_{pll} = \omega_0$, and $\omega = 0$.

Further, after reorganizing (1), we obtain the following second-order ordinary differential equation (ODE):

$$\begin{cases} \dot{\varphi} = \omega \\ M\dot{\omega} = P_m - P_e - \alpha \cos\varphi\omega \end{cases} \tag{2}$$

where $M = 1/k_{ipll}, P_m = i_d X_g, P_e = U_g \sin\varphi$, and $\alpha = k_{ppll} U_g / k_{ipll}$. Here $M$, $P_m$, and $P_e$ represent the equivalent inertial, mechanical torque (force), and electromagnetic torque (force), respectively. Note that $P_m$ is a function of $i_d$, and $P_e$ is a function of $U_g$, and they are very important in the following analysis. As equation (2) also describes the mass-particle motion under imbalanced forces and is similar to the swing equation (SE) of SG for the rotor's motion, we call it the generalized swing equation (GSE)[42-44]. Although it has been used in the during-fault stage-2 staility analysis of converter systems, we will see that it plays a dominant role in the RDPS as well.

Next, we find that the stage-3 DAE model can be divided into a driving subsystem (mainly including the PLL and the active current climbing control)

$$\begin{cases} \dot{\varphi} = \omega \\ \dot{\omega} = k_{ppll} \dot{u}_{tq} + k_{ipll} u_{tq} \\ \dot{i}_d = K_{ramp} \\ u_{tq} = i_d X_g - U_g \sin\varphi \end{cases} \tag{3}$$

and a response one (mainly including the TVC)

$$\begin{cases} \dot{i}_q = K_{pV} \dot{U}_t + K_{iV} (U_t - U_{tref}) \\ u_{td} = -i_q X_g + U_g \cos\varphi \\ U_t = \sqrt{u_{td}^2 + u_{tq}^2} \end{cases} \tag{4}$$

Here $K_{ramp}$ denotes the climbing rate, and $i_{d3}$ should increase from $i_{d2}$ to a recovery value $i_{d4}$ ($i_{d4} = i_{d1}$).

Under the quasi-steady-state assumption, the driving system is dominant and is still a GSE system, with an extremely slowly increasing $i_d$. We find then the classical "fish-like" basin boundary of the GSE under different $i_d$'s (including $i_{d3} = i_{d2}$ at the beginning of stage 3 and $i_{d3} = i_{d4} = i_{d1}$ at the end of stage 3) and the local dynamics near the unstable equilibrium point, as shown in Fig. 3. Clearly the basin of attraction moves to the lower-left part with an extremely slow speed, to be compared with the fast-moving trajectory. Therefore, if the stage-3 initial state (open triangle) is within its basin (blue region), it will be stable. Otherwise, it will move to the upper-right and becomes unstable. This indicates that the system fate for whether it is stable or unstable after all 4 stages has been already completely determined at the beginning of stage 3 under $i_{d3} = i_{d2}$, and the following dynamical behavior of stages 3 and 4 is unimportant and does not be considered.

3) Therefore, for the RDPS transient stability analysis, we can establish a unified model based on the same GSE (2) under the parameter variations, including $U_g$ (as an external parameter) and $i_d$ (as an internal parameter), i.e., $U_{g1}$ and $i_{d1}$ for stage 1, $U_{g2}$ and $i_{d2}$ for stage 2, and $U_{g3}$ and $i_{d3}$ ($i_{d3} = i_{d2}$ virtually) for stage 3, which just correspond to the pre-fault, during-fault, and post-fault three stages in the traditional power system, respectively. Usually $U_{g3} = U_{g1} = 1.0$ are chosen and $i_{d1}$ is determined by the stable equilibrium point of stage 1. There are only two tunable parameters: $U_{g2}$ and $i_{d2}$. This unified model, which has treated diversified wind-and-solar models in a unified manner, has been added in the last line in Table 1 and will construct a theoretical basis for further analysis.

## Transient synchronization stability analysis

The critical clearing angle (CCA) for the maximal deviation angle and the critical clearing time (CCT) for the maximal tolerable time of during-fault stage 2 are very important. They provide a good guide for operation and control and work as a perfect index for a theory test.

1) As the first step, we treat $\alpha = 0$ in the GSE (2) for a conservative system, which can be viewed as a mechanical system, i.e., a forced mass point under the-same-form potential but different parameters of the pre-fault (stage 1), during-fault (stage 2), and post-fault (stage 3). Based on the above unified model, in the whole stage 3, $i_{d3} = i_{d2}$ and $U_{g3} = U_{g1} = 1.0$ are treated. The total energy should include kinetical energy and potential energy, which further includes a position potential ($-P_m\varphi$) and an electromagnetic potential ($-U_g \cos\varphi$)[6].

The classical equal-area criterion (EAC)[6] widely used in the traditional transient stability analysis will be extended here first. Now the accelerated area $E_{ac} = \int_{\varphi_1}^{\varphi_{cr}^{[1]}} (P_{m2} - U_{g2}\sin\varphi)d\varphi$ denotes the release of the potential energy or, equivalently, the increase of kinetic energy at stage 2. Here $\varphi_1 = \arcsin(P_{m1}/U_{g1}) = \arcsin(i_{d1}X_g/U_{g1})$ is for the stable equilibrium point of stage 1, which is already known, and $\varphi_{cr}^{[1]}$ denotes the CCA at this first approximation. Meanwhile, the decelerated area $E_{de} = \int_{\varphi_{cr}^{[1]}}^{\varphi_{3u}} (\sin\varphi - P_{m2})d\varphi$ denotes the increase of the potential energy or, equivalently, the decrease of kinetic energy at stage 3. Here $\varphi_{3u} = \pi - \arcsin(P_{m3}/U_{g3}) = \pi - \arcsin(i_{d2}X_g/U_{g1})$ denotes the unstable equilibrium point of stage 3 under $i_{d3} = i_{d2}$ and $U_{g3} = U_{g1} = 1.0$, since for a stable system, $\varphi$ larger than $\varphi_{3u}$ is not allowed. Therefore, based on $E_{ac} = E_{de}$, we have the explicit expression for $\varphi_{cr}^{[1]}$:

$$\varphi_{cr}^{[1]} = \arccos\frac{U_{g2}\cos\varphi_1 - \cos\varphi_{3u} + P_{m2}(\varphi_1 - \varphi_{3u})}{U_{g2} - 1} \tag{5}$$

The trajectory on the position-velocity ($\varphi$-$\omega$) plane and the corresponding EAC analysis are schematically shown in Figs. 4a and 4b, respectively.

2) Then we will consider the effect of a velocity jump, which is apparent in Fig. 3. It may bring an additional change of kinetical energy from stages 1 to 2, and also from stages 2 to 3. After considering this new effect and still on the basis of the energy conservation principle, we get the explicit nonlinear algebraic equation for $\varphi_{cr}^{[2]}$ (see Methods for details). In a contrast, the trajectory and the corresponding EAC analysis are schematically shown in Figs. 4c and 4d, respectively. Now $\varphi_{cr}^{[2]}$ slightly shifts and $E_{ac} \neq E_{de}$.

3) As the third improved approximation, we consider the $\alpha \neq 0$ damping in the GSE (2). Usually there is no explicit solution. However, it is available for zero $\alpha$. Therefore, relying on the explicit form of the trajectory under $\alpha = 0$ and the additive dissipative energy due to the damping, we obtain the explicit form of $\varphi_{cr}^{[3]}$, based on the already-known $\varphi_{cr}^{[2]}$ (see Methods for details). In contrast, the trajectory and the corresponding EAC analysis are schematically shown in Figs. 4e and 4f, respectively.

## Simulation results

To verify the theory, we establish a detailed single PMSG grid-connected system with the switching control strategies in Matlab/Simulink (Fig. 1c). Multiple EMT simulations are conducted under different voltage drops $U_{g2}$ and active current references $i_{d2}$. The corresponding CCA is recorded and regarded as the real value. Meanwhile, the three CCAs under the above progressive approximations, e.g., $\varphi_{cr}^{[1]}$, $\varphi_{cr}^{[2]}$, and $\varphi_{cr}^{[3]}$, are calculated by equations (5), (12), and (17), respectively. After getting the CCA, the corresponding CCT compared with the fault trajectory, can be yielded. The results for their CCA's and CCT's are summarized in Tables 2a and 2b. From these comparisons, we mainly find that the error of $\varphi_{cr}^{[1]}$ is the largest, around 10%, and the error of $\varphi_{cr}^{[2]}$ is the next, reduced to around 4%, and the error of $\varphi_{cr}^{[3]}$ is the smallest, around only 1%, in terms of absolute values for different cases. Extensive simulations are conducted and the same accuracies are found. The detailed results are given in Supplementary Note II and the additional verifications on SpaceR real-time simulation platform are presented in Supplementary Note III.

In addition, a PMSG field consisting of $n$ (e.g., $n$=10) PMSG systems is studied. After using the classical single-machine-multiplication technique[45,46], the PMSG field can be aggregated into a single PMSG system, as usually the renewable devices in a field show a very high consistency and they go into the LVRT simultaneously. Correspondingly, the CCT results are summarized in Table 2c. See Supplementary Note IV for more details. The original stage-1 EMT model is a ($14n$+2=142) order DAE system, including a 14th-order DAE for each PMSG plus a second-order ODE for transmission line dynamics. It is amazing to discover that this complicated, switching, high-dimensional nonlinear system can still be perfectly analyzed. Again, except the computational accuracy of only 1% error in average, the computational efficiency is extremely high, as it nearly does not cost any time.

In conclusion, similar to unified theories in mathematics or physics, for the first time a unified wind-and-solar-energy GSE transient model and an improved EAC method are proposed, to understand and analyze the bulk transient behaviors of diversified renewable devices in a unified manner. Compared to the traditional SG-dominated power system, well-known as the largest man-made complex system, the emerging RDPS dynamics becomes much more complicated. However, to our surprise, by catching the dominant switching dynamical characteristics under the LVRT, this spatial-temporal large-scale complex system can be well described by a second-order nonlinear GSE and analyzed by an improved EAC method, exhibiting only 1% error with respect to the detailed EMT simulations. This unified theory not only manifests the power of nonlinear dynamical system analysis, but also lays a solid foundation to electrical power engineering! A larger combined renewable-and-SG-system stability remains to be investigated.

# Methods

## Transient synchronization stability analysis

1) Based on the unified model including the three stages 1, 2, and 3, the classical EAC method can be used, to obtain the first approximation of the CCA, $\varphi_{cr}^{[1]}$.

2) For the second approximation, we should consider the effects of the velocity jump and the associated change of kinetical energy at each switch. In this effect, from stages 1 to 2, as $U_{g1}$ jumps to $U_{g2}$, and $i_{d1}$ jumps to $i_{d2}$ instantaneously, $u_{tq}$ also jumps according to (1), i.e.,

$$\Delta u_{tq} = (i_{d2} - i_{d1})X_g - (U_{g2} - 1)\sin\varphi_1 \tag{6}$$

By the influence of $u_{tq}$, $\omega$ should also jump from 0 to $\omega_1$, according to (1) ;

$$\omega_1 = k_{ppll}\Delta u_{tq} = k_{ppll}[(i_{d2} - i_{d1})X_g - (U_{g2} - 1)\sin\varphi_1] \tag{7}$$

Similarly, from stage 2 to stage 3, we have

$$\omega_3 - \omega_2 = k_{ppll}(U_{g2} - 1)\sin\varphi_{cr}^{[2]} \tag{8}$$

Here $\varphi_{cr}^{[2]}$ denotes the CCA at the second approximation.

By the energy conservation principle, we get

$$-\frac{1}{2}M\omega_1^2 + \frac{1}{2}M(\omega_2^2 - \omega_3^2) = \int_{\varphi_1}^{\varphi_{cr}^{[2]}} (P_{m2} - U_{g2}\sin\varphi)d\varphi + \int_{\varphi_{cr}^{[2]}}^{\varphi_{3u}} (P_{m2} - \sin\varphi)d\varphi \tag{9}$$

Note that the left term vanishes in the first approximation. In addition, we have the (constant) total energy of stage 2, $h_2 = \frac{1}{2}M\omega_1^2 - P_{m2}\varphi_1 - U_{g2}\cos\varphi_1$ and further the velocity

$$\omega = \sqrt{2(P_{m2}\varphi + U_{g2}\cos\varphi + h_2)/M} \tag{10}$$

for the corresponding curve 2 in Fig. 4c. Similarly, we have the (constant) total energy of stage 3, $h_3 = -P_{m2}\varphi_{3u} - \cos\varphi_{3u}$ and further the velocity

$$\omega = \sqrt{2(P_{m2}\varphi + \cos\varphi + h_3)/M} \tag{11}$$

for the corresponding curve 3 in Fig. 4c.

They are restricted by equation (8). Therefore, we obtain the nonlinear algebraic equation for $\varphi_{cr}^{[2]}$:

$$\sqrt{2(P_{m2}\varphi_{cr}^{[2]} + \cos\varphi_{cr}^{[2]} + h_3)/M} - \sqrt{2(P_{m2}\varphi_{cr}^{[2]} + U_{g2}\cos\varphi_{cr}^{[2]} + h_2)/M} = k_{ppll}(U_{g2} - 1)\sin\varphi_{cr}^{[2]} \tag{12}$$

which can be easily solved. In a contrast, the trajectory and the corresponding EAC analysis are schematically shown in Figs. 4c and 4d, respectively. Now $\varphi_{cr}^{[2]}$ slightly shifts and $E_{ac} \neq E_{de}$.

3) As the third improved approximation, we should consider the nonzero $\alpha$ damping in the GSE (2). Then the energy conservation relation becomes

$$-\frac{1}{2}M\omega_1^2 + \frac{1}{2}M(\omega_2^2 - \omega_3^2) = \int_{\varphi_1}^{\varphi_{cr}^{[3]}} (P_{m2} - U_{g2}\sin\varphi - \alpha\cos\varphi\omega)d\varphi + \int_{\varphi_{cr}^{[3]}}^{\varphi_{3u}} (P_{m2} - \sin\varphi - \alpha\cos\varphi\omega)d\varphi \tag{13}$$

where the dissipative energy due to the damping has been incorporated and $\varphi_{cr}^{[3]}$ denotes the CCA at the third approximation. Let $S_d$ denote the contribution due to damping:

$$S_d = \int_{\varphi_1}^{\varphi_{cr}^{[3]}} (\alpha \cos\varphi\omega) d\varphi + \int_{\varphi_{cr}^{[3]}}^{\varphi_{3u}} (\alpha \cos\varphi\omega) d\varphi \tag{14}$$

As $S_d$ depends on the time-varying velocity, usually it cannot be explicitly obtained. In this approximation, we use the conservative-system solution curves in equations (10) and (11) for the second approximation and yield

$$S_d \approx \int_{\varphi_1}^{\varphi_{cr}^{[2]}} (\alpha \cos\varphi \sqrt{2(P_{m2}\varphi + U_{g2}\cos\varphi + h_2)/M}) d\varphi + \int_{\varphi_{cr}^{[2]}}^{\varphi_{3u}} (\alpha \cos\varphi \sqrt{2(P_{m2}\varphi + \cos\varphi + h_3)/M}) d\varphi \tag{15}$$

Then, comparing equations (13) and (9), we have

$$(U_{g2} - 1)(\cos\varphi_{cr}^{[3]} - \cos\varphi_{cr}^{[2]}) = S_d \tag{16}$$

and further

$$\varphi_{cr}^{[3]} = \arccos(\frac{S_d}{U_{g2} - 1} + \cos\varphi_{cr}^{[2]}) \tag{17}$$

based on the already-known $\varphi_{cr}^{[2]}$. In a contrast, the trajectory and the corresponding EAC analysis are schematically shown in Figs. 4e and 4f, respectively.

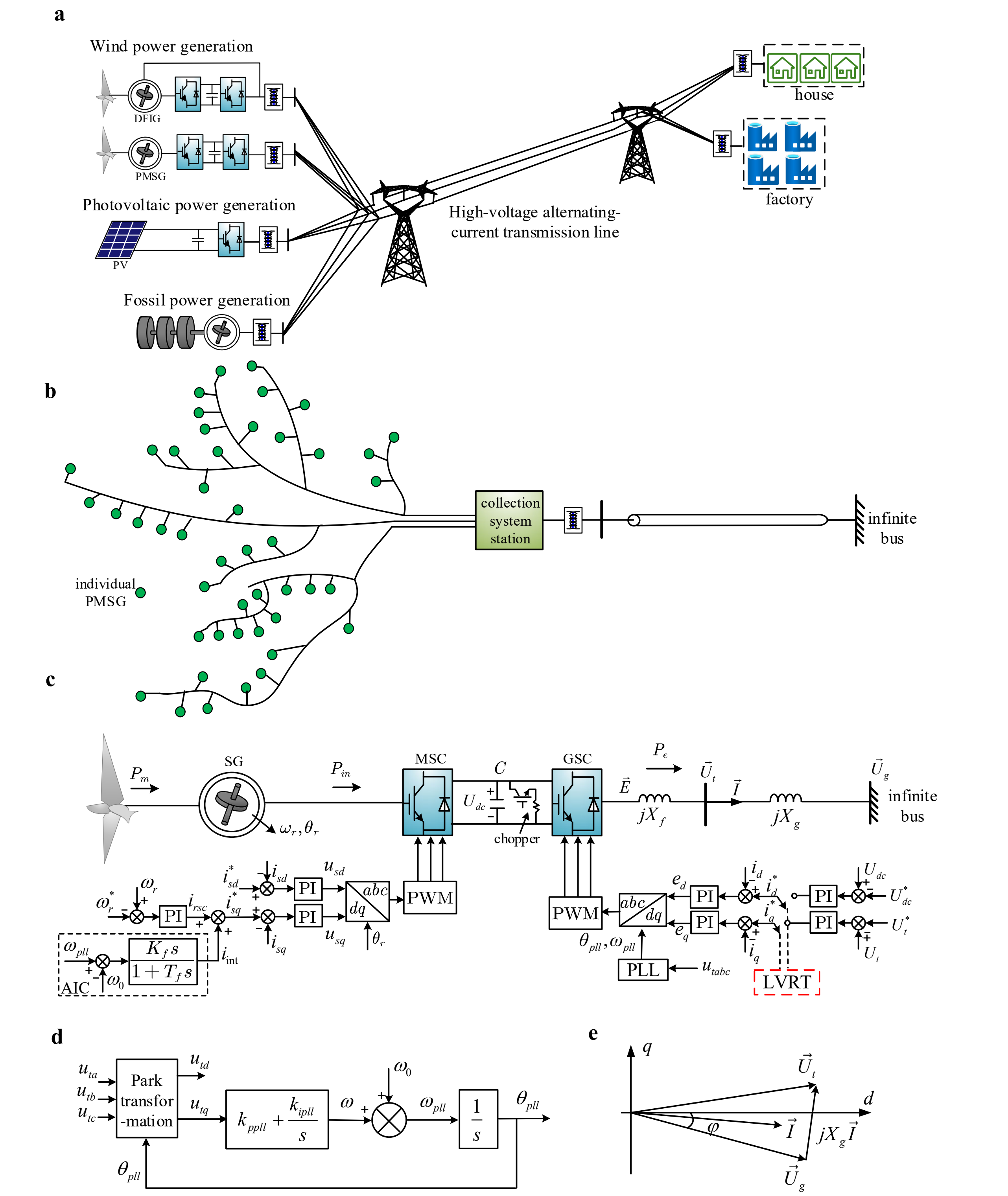

**Fig. 1 | Schematical show of a renewable dominated power system. a**, The renewable energy generations mainly including DFIG and PMSG for wind power and PV for solar power are connected to the AC power grid by power electronic converters, accompanying with the fossil power generation in the traditional power system. **b**, Hundreds of wind turbines are interconnected within a wind farm. **c**, A single PMSG infinite-bus system, to show the detailed control structure. **d**, The PLL control structure used in the PMSG system. **e**, The voltage phasor diagram.

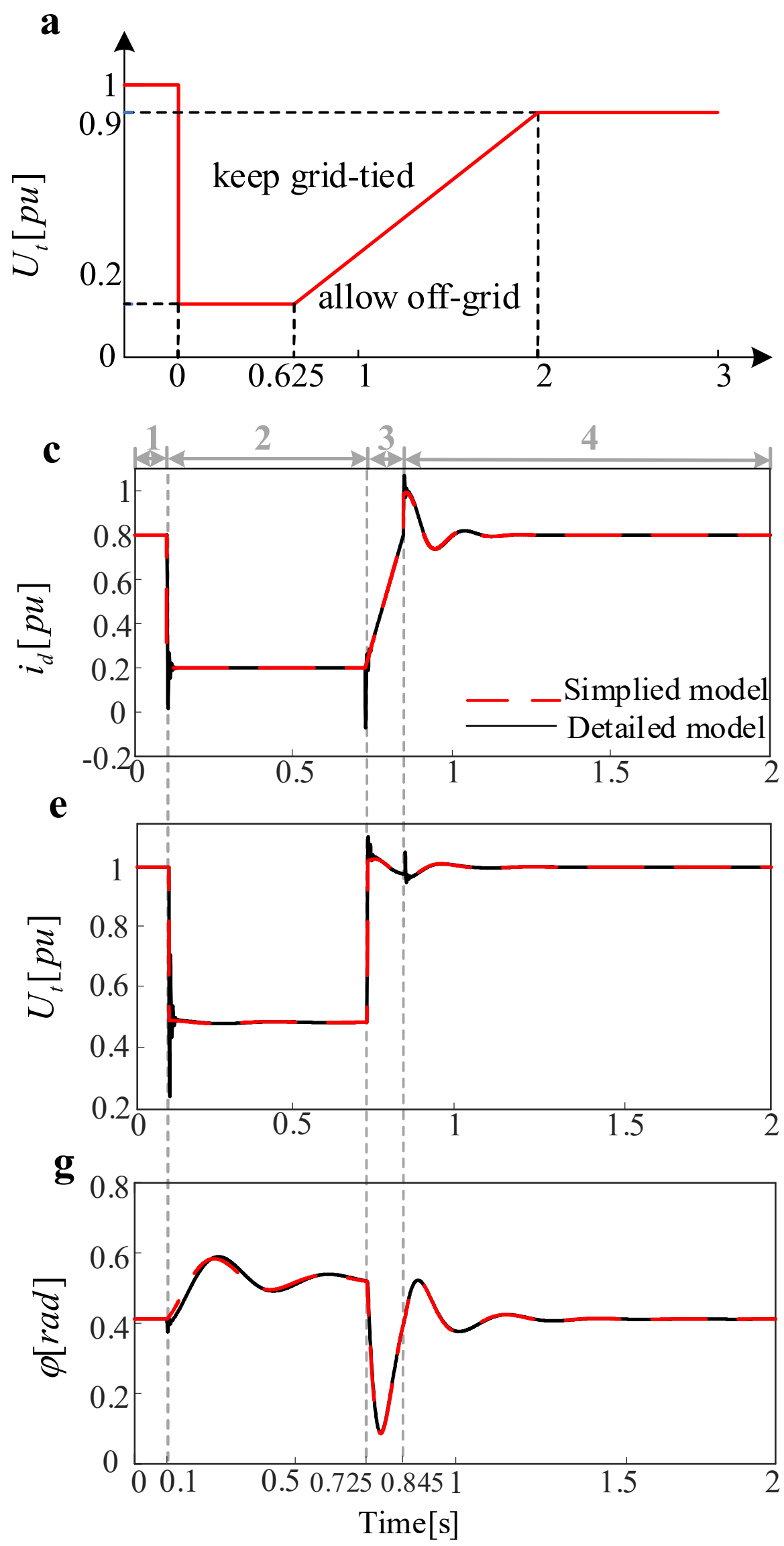

**Fig. 2 | Low voltage ride-through curve and model verification of a PMSG grid-tied system**. **a,** The LVRT curve of wind turbines, according to the Chinese grid code. When the terminal voltage $U_t$ drops below 0.8 pu, the wind turbine has to switch to the LVRT control. Below the red line, the wind turbine can be off-grid. **b** Time series of $U_g$ in the PMSG infinite-bus system, under $U_g$ drops sharply from an initial $U_g$ = 1 pu to 0.2 pu at 0.1 s, and then recovers to 1 at 0.725 s. Four stages 1, 2, 3, and 4 are shown. **c-h**, Comparisons of time series of $i_d$, $i_q$, $U_t$, $U_{dc}$, $\varphi$, and $\omega$, respectively, to show the similarity between the simplified switching model (red dotted line) and the detailed EMT model (black solid line).

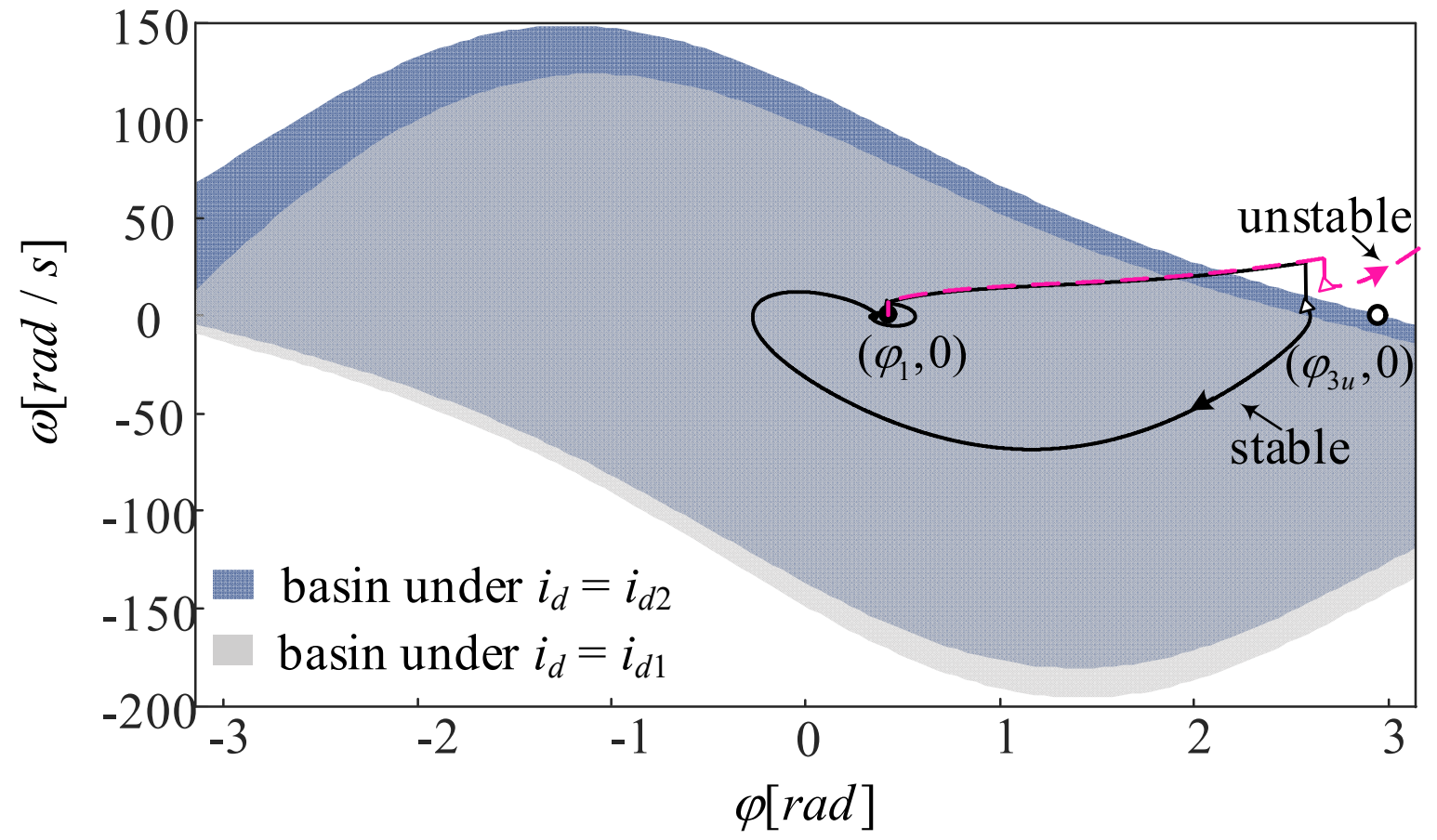

**Fig. 3 | "Fish-like" basin of attraction of the GSE under different $i_d$'s and comparison of stable and unstable trajectories.** Under the quasi-steady-state assumption, the stage-3 system can be viewed as a series of GSE under different $i_d$'s, and their corresponding basins move very slowly. The blue area represents the basin of attraction of the GSE under $i_d = i_{d2} = 0.4$ pu at the beginning of stage 3, whereas the gray area represents that under $i_d = i_{d1} = 0.8$ pu at the end of stage 3. The triangle represents the system state at the beginning of stage 3, when $U_g$ is recovered. ($\varphi_1$, 0) denotes the initial stable equilibrium point of stage 1 (solid circle), and ($\varphi_{3u}$, 0) the unstable equilibrium point of stage 3 (open circle). Clearly, if the system state is within the blue area, the trajectory will converge and become stable finally. Otherwise, the trajectory will diverge and become unstable. Here the frequency jump at the beginning of stage 3 is clear. Therefore, the system fate is solely determined by the system state at the beginning of stage 3, and this character helps simplify transient synchronization stability analysis of the RDPS significantly.

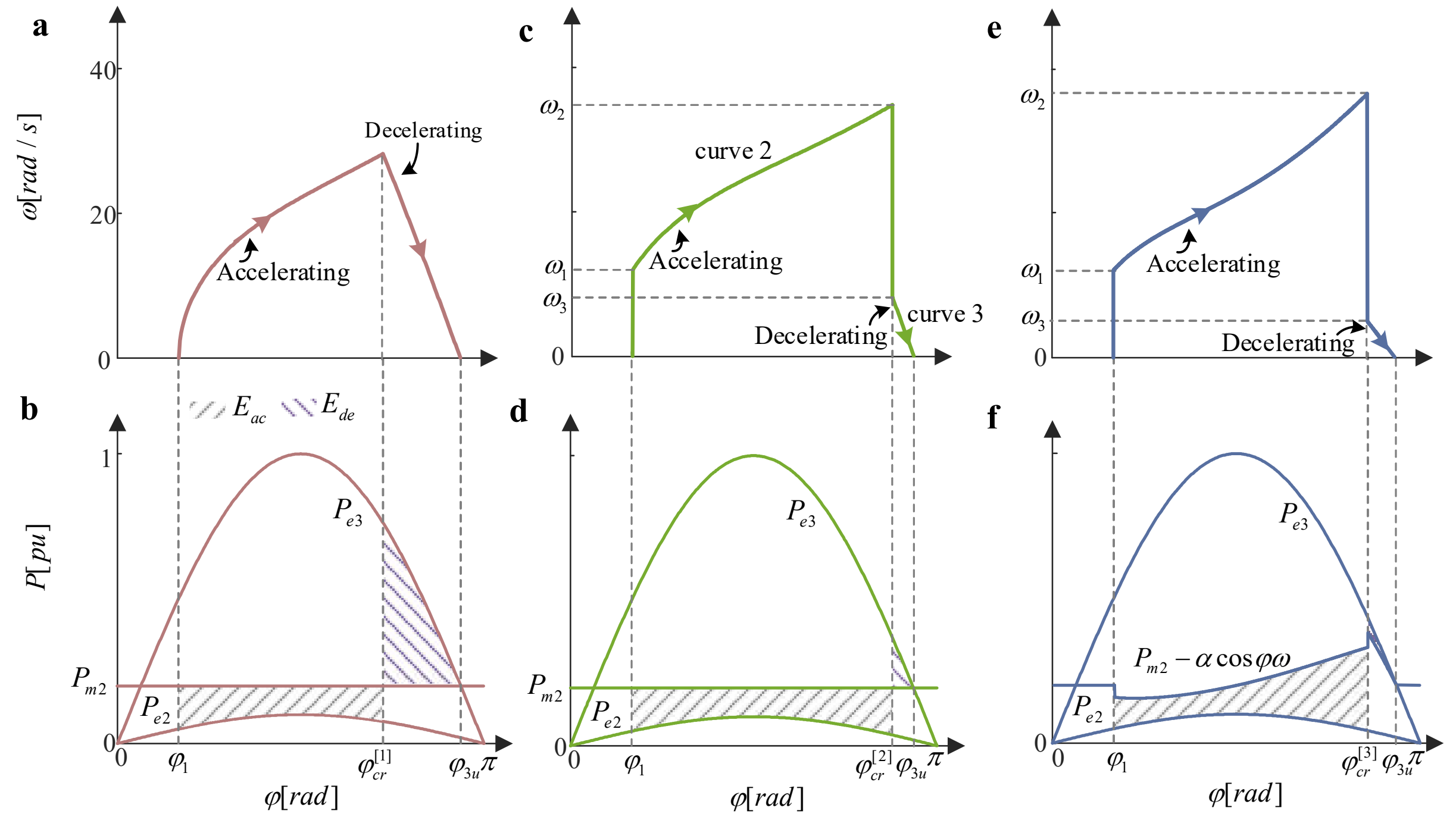

**Fig. 4 | Comparison of trajectory on the position-velocity plane and the corresponding EAC analysis for the three progressive approximations**. **a-b**, The first approximation result of $\varphi_{cr}^{[1]}$, based on that $\alpha = 0$ and $E_{ac} = E_{de}$ (i.e., the accelerated area $E_{ac}$ equals the decelerated area $E_{de}$). **c-d**, The second approximation result of $\varphi_{cr}^{[2]}$, based on that $\alpha = 0$ and incorporating the additive kinetic energy inputs due to one discontinuous velocity jump from 0 to $\omega_1$ at $\varphi_1$, and the other from $\omega_2$ to $\omega_3$ at $\varphi_{cr}^{[2]}$. **e-f**, The third approximation result of $\varphi_{cr}^{[3]}$, based on that $\alpha \neq 0$ and relying on the already-obtained $\varphi_{cr}^{[2]}$ from the conservative system analysis.

**Table 1 Summary of four-stage models for typical renewable grid-tied systems and unified model**

| Renewable grid-tied system | Four-stage models | | | |
|---|---|---|---|---|
| | Stage 1 | Stage 2 | Stage 3 | Stage 4 |
| PMSG without AIC | Fifth-order DAE (S11) | Second-order GSE (S14) | Fourth-order DAE (S17) | Fifth-order DAE (S11) |
| PMSG with AIC | Fifth-order DAE (S25) | Second-order GSE (S14) | Fourth-order DAE (S17) | Fifth-order DAE (S25) |
| PV | Fifth-order DAE (S28) | Second-order GSE (S14) | Fourth-order DAE (S17) | Fifth-order DAE (S28) |
| DFIG | Fifth-order DAE (S45) | Second-order GSE (S48) | Fourth-order DAE (S52) | Fifth-order DAE (S45) |
| Unified model | GSE ($U_g = U_{g1}$, $i_d = i_{d1}$) | GSE ($U_g = U_{g2}$, $i_d = i_{d2}$) | GSE ($U_g = U_{g1}$, $i_d = i_{d2}$) | |

The four-stage models for all typical renewable grid-tied systems including the PMSG without AIC, PMSG with AIC, PV, and DFIG, their stage-1 and stage-4 models established. They are completely different, due to different control structures and topologies, but their stage-2 and stage-3 models are similar. For each case, its stage-1 and stage-4 models are identical. In addition, the DFIG system models are special. By considering the four scaled parameters, the DFIG stages-2 and stage-3 models are similar to the other cases. After establishing the four-stage models for all typical renewable grid-tied systems and considering the dynamical characteristics of stage 3, all of them in a whole can be unified into the unified model including the three stages 1, 2, and 3 only, as shown at the last line. Now, in the transient stability analysis $i_{d3} = i_{d2}$ should be considered for the whole stage 3.

**Table 2 Comparison of CCA and CCT results and error analysis**

**a**

| $U_{g2}$ | $i_{d2}$ | $\varphi_{cr}$ | $\varphi_{cr}^{[1]}$ | error | $\varphi_{cr}^{[2]}$ | error | $\varphi_{cr}^{[3]}$ | error |
|---|---|---|---|---|---|---|---|---|
| 0.2 | 0.4 | 2.739 | 2.562 | -6.46% | 2.825 | 3.14% | 2.676 | -2.30% |
| | 0.5 | 2.595 | 2.349 | -10.25% | 2.705 | 4.23% | 2.540 | -2.11% |
| 0.1 | 0.25 | 2.769 | 2.591 | -6.43% | 2.870 | 3.65% | 2.735 | -1.22% |
| | 0.4 | 2.600 | 2.276 | -12.46% | 2.711 | 4.27% | 2.565 | -1.35% |
| 0 | 0.2 | 2.726 | 2.409 | -11.63% | 2.808 | 3.01% | 2.710 | -0.59% |
| | 0.3 | 2.623 | 2.238 | -14.68% | 2.721 | 3.74% | 2.616 | -0.27% |
| Average | | | | 10.32% | | 3.67% | | 1.31% |

**b**

| $U_{g2}$ | $i_{d2}$ | $t_{cr}$ | $t_{cr}^{[1]}$ | error | $t_{cr}^{[2]}$ | error | $t_{cr}^{[3]}$ | error |
|---|---|---|---|---|---|---|---|---|
| 0.2 | 0.4 | 0.166 | 0.159 | -4.22% | 0.169 | 1.81% | 0.164 | -1.20% |
| | 0.5 | 0.102 | 0.095 | -6.87% | 0.105 | 2.94% | 0.100 | -1.96% |
| 0.1 | 0.25 | 0.163 | 0.156 | -4.29% | 0.167 | 2.45% | 0.162 | -0.61% |
| | 0.4 | 0.090 | 0.081 | -10.00% | 0.092 | 2.22% | 0.089 | -1.11% |
| 0 | 0.2 | 0.112 | 0.103 | -8.04% | 0.114 | 1.79% | 0.111 | -0.89% |
| | 0.3 | 0.083 | 0.074 | -10.84% | 0.085 | 2.41% | 0.083 | 0.00% |
| Average | | | | 7.38% | | 2.27% | | 0.96% |

**c**

| $U_{g2}$ | $i_{d2}$ | $t_{cr}$ | $t_{cr}^{[1]}$ | error | $t_{cr}^{[2]}$ | error | $t_{cr}^{[3]}$ | error |
|---|---|---|---|---|---|---|---|---|
| 0.1 | 0.1 | 0.229 | 0.221 | -3.49% | 0.232 | 1.31% | 0.227 | -0.87% |
| | 0.2 | 0.118 | 0.108 | -8.47% | 0.120 | 1.69% | 0.116 | -1.69% |
| 0.05 | 0.1 | 0.133 | 0.124 | -6.77% | 0.135 | 1.50% | 0.131 | -1.50% |
| | 0.2 | 0.093 | 0.086 | -7.53% | 0.097 | 4.30% | 0.093 | 0.00% |
| 0 | 0.1 | 0.100 | 0.090 | -10.00% | 0.102 | 2.00% | 0.099 | -1.00% |
| | 0.2 | 0.078 | 0.068 | -12.82% | 0.081 | 3.85% | 0.078 | 0.00% |
| Average | | | | 8.18% | | 2.44% | | 0.84% |

**a,** Comparison of CCA results under EMT simulation in Simulink, including $\varphi_{cr}$, and those by three approximations in the improved EAC analysis, $\varphi_{cr}^{[1]}$, $\varphi_{cr}^{[2]}$, and $\varphi_{cr}^{[3]}$, calculated by equations (5), (12), and (17), respectively. Here $U_{g2}$ and $i_{d2}$ are in pu, and $\varphi_{cr}$, $\varphi_{cr}^{[1]}$, $\varphi_{cr}^{[2]}$, and $\varphi_{cr}^{[3]}$ are in *rad*. In average of the absolute values, the error of $\varphi_{cr}^{[1]}$ is around 10%, the error of $\varphi_{cr}^{[2]}$ is approximately 4%, and the error of $\varphi_{cr}^{[3]}$ is the smallest, approximately 1%. In the test, the wind fluctuation is not considered because it has much longer time, compared to the transient response time, the infinite bus is replaced by a three-phase programmable voltage source, and the voltage source amplitude is changed to simulate the infinite-bus voltage sag fault. **b**, Comparison of CCT results. Here $t_{cr}$, $t_{cr}^{[1]}$, $t_{cr}^{[2]}$, and $t_{cr}^{[3]}$ are in second. Again, the error of $t_{cr}^{[3]}$ is the smallest, approximately around 1%. **c**, Comparison of CCTs of a wind farm consisting of 10 wind turbines. After using the classical single-machine multiplication method, the CCTs under the three approximations show the similar results. See the supplementary material for more details.